\documentstyle{article}

\font\bb=msym10
\newcommand{\cn}{\mbox{\bb C}}
\newcommand{\rn}{\mbox{\bb R}}

\begin{document}
\setlength{\baselineskip}{7mm}
\bibliographystyle{plain}

\title{Quantum determinants}
\author{ \it Ulrich Meyer  \\
{\it DAMTP, University of Cambridge}\\
{\it  Cambridge CB3 9EW, England} \\
{\small\it e-mail U.Meyer@amtp.cam.ac.uk}}
\date{\it  June 1994}
\maketitle
\vspace{-8cm} \hfill DAMTP/94-54 \vspace{8cm}

\begin{abstract}
We show how to construct
central and   grouplike quantum determinants
for
FRT algebras $A(R)$. As an application of the
general construction
we give a quantum determinant for the
q-Lorentz
group.
\end{abstract}

\section{Introduction}

FRT algebras $A(R)$ were introduced in
\cite{resh/90}
as the associative $\cn$-algebras
generated by
1 and $t^{a}_{\;b};\; a,b=1,\ldots ,n$
with relations
\begin{equation}\label{arrels}
R^{ab}_{\;cd}t^{c}_{\;e}t^{d}_{\;f}=
t^{b}_{\;d}t^{b}_{\;c}R^{cd}_{\;ef},
\end{equation}
where $R$ is  an
invertible solution of
the $n$-dimensional
matrix quantum Yang-Baxter equation.
These algebras are   non-commutative
generalisations of
the algebras of polynomial functions
on $n\times n$-matrices and
play the r\^{o}le
of q-matrix symmetries for q-deformed
systems
\cite{DAMTP/92-12}.
They are bialgebras with coproduct
$\Delta t^{a}_{\;b} =
t^{a}_{\;c}\otimes t^{c}_{\;b}$
and counit $\epsilon (t^{a}_{\;b})=
\delta^{a}_{\;b}$,
and have so-called
{\em fundamental representations}
$\rho_{\pm}$
in the algebra of
$n\times n$-matrices given by
\cite{majid/4/89}
\begin{equation}\label{fundrep}
\rho_{+}(t^{a}_{\;c})^{b}_{\;d}=
R^{ab}_{\;cd},\;\;\;\;
\rho_{-}(t^{a}_{\;c})^{b}_{\;d}=
R^{-1ba}_{\;\;\;\;\;\;dc}.
\end{equation}
The fact that $\rho_{\pm}$ extend
as algebra
maps has the consequence that $A(R)$
is {\em dual
quasitriangular} (see \cite{majid/4/89}
for details).

In order
to obtain generalisations of unitary matrix
groups one often   divides by a further
`metric relation'
of the form
\begin{equation}\label{metrel}
t^{a}_{\;c}t^{b}_{\;d}g^{cd}=g^{ab},
\end{equation}
and the quotient $A$ of $A(R)$ by the ideal
generated by this relation
is then
a Hopf algebra with obvious antipode.
However, for mathematical reasons, one
would like the
quotient $A$ also to be  dual
quasitriangular. This is the case
if the fundamental
representations $\rho_{\pm}$
descend to a representation
of $A$, i.e. if they respect
(\ref{metrel}). It is easily
seen that this requirement
is satisfied iff
\begin{equation}\label{qgrpnorm}
R^{ae}_{\;cm}R^{bm}_{\;df} g^{cd}
=g^{ab}\delta^{e}_{\;f},\;\;\;\;
R^{-1\;ea}_{\;\;\;\;\;\;mc}
R^{-1\;mb}_{\;\;\;\;\;\;fd}g^{cd}
=g^{ab}\delta^{e}_{\;f}.
\end{equation}
These relations place a
restriction on the choice
of the metric $g$.

Generalisations
of special  matrix
groups (i.e. matrices with
determinant one), are obtained by
dividing $A$ by a q-deformed determinant
relation. Such relations
are known for a number of examples, but a
general construction
seems to be missing so far. In this
paper, we provide
such a construction and show that the
resulting
quantum determinants ${\cal D}$
 are central and {\em grouplike}, i.e.
`multiplicative': $\Delta {\cal D}=
{\cal D}\otimes {\cal D}$.

\section{Quantum determinants}

In order to construct a q-determinant
in $A(R)$,
we need to
find a $\cn$-valued tensor
$\varepsilon_{a_{1}\ldots a_{n}}$
which is completely {\em q-antisymmetric}
in the
sense that it solves the equation
\begin{equation}\label{quanti}
\varepsilon_{a_{1}\ldots a_{k}a_{k+1}
\ldots a_{n}}
= -\lambda\varepsilon_{a_{1}\ldots ij
\ldots a_{n}}
R^{j\;i}_{\;a_{k}a_{k+1}}
\end{equation}
for any two adjacent indices $a_{k}$ and
$a_{k+1}$.
Here $\lambda$ is a suitable normalisation
factor.
Solutions to these equations can
be easily calculated,
and in well-behaved cases  the
space of solutions
will be  one-dimensional.

There is also a general R-matrix formula
for such epsilon tensors  \cite{DAMTP/94-20}.
The setting in this paper was that $A(R)$
acted on an algebra of q-forms
given by a second
R-matrix $R^{'}$, which has to
obey certain relations
with $R$. This space of forms was assumed
to have a unique top form. An epsilon
tensor was then
constructed from this top form by
differentiation and is
given in terms of the matrix $R^{'}$.
However, this
R-matrix formula is not very
useful for actually
calculating the epsilon tensor.
Checking the assumption that
there is a unique
top form is essentially tantamount
to verifying that
(\ref{quanti}) has a one-dimensional
space of solutions.
This already gives the epsilon tensor and
thus there is no need to use the rather
complicated R-matrix formula. Moreover, in our
approach there is no need to find the
second R-matrix $R^{'}$.

For the following, we assume that there is
a unique solution
$\varepsilon_{a_{1}\ldots a_{n}}$
of (\ref{quanti}). In terms of
this epsilon tensor we
define the {\it quantum determinant}
$\cal D$
in $A(R)$ as
$${\cal D}=\nu^{-1}
\varepsilon_{a_{1}\ldots a_{n}}
t^{a_{1}}_{\;b_{1}}
\ldots
t^{a_{n}}_{\;b_{n}}
\varepsilon^{b_{n}\ldots b_{1}},$$
where the normalisation factor $\nu$ is given
by
$$
\nu =
\varepsilon_{a_{1}\ldots a_{n}}
\varepsilon^{a_{n}\ldots a_{1}},
$$
and the
epsilon tensor
with upper indices is defined as
\begin{equation}\label{epup}
\varepsilon^{a_{1}\ldots a_{n}}
=
\varepsilon_{b_{1}\ldots b_{n}}
g^{a_{1}b_{1}}
\ldots
g^{a_{n}b_{n}}
\end{equation}

{\bf Proposition.}
\it
The q-determinant ${\cal D}$
is central and grouplike
and the
 generators of $A(R)$ obey the relation
\begin{equation}\label{detrel}
\varepsilon_{a_{1}\ldots a_{n}}
t^{a_{1}}_{\;b_{1}}
\ldots
t^{a_{n}}_{\;b_{n}}
={\cal D} \varepsilon_{b_{1}\ldots b_{n}}.
\end{equation}
In the quotient $A$ of $A(R)$,
one also finds
${\cal D}^{2}=1$
for the square of the q-determinant.
\rm

{\bf Proof.}
By virtue of the relations
(\ref{arrels}) and the
properties of the epsilon tensor,
the element
$\varepsilon_{a_{1}\ldots a_{n}}
t^{a_{1}}_{\;b_{1}}
\ldots
t^{a_{n}}_{\;b_{n}}$
is a solution of  (\ref{quanti}) and
thus by assumption
eigenvector of the
one-dimensional projector
$P^{a_{1}\ldots a_{n}}_{\;b_{1}
\ldots b_{n}}=\nu^{-1}
\varepsilon^{a_{n}\ldots a_{1}}
\varepsilon_{b_{1}\ldots b_{n}}$, i.e.
$$\varepsilon_{a_{1}\ldots a_{n}}
t^{a_{1}}_{\;b_{1}}
\ldots
t^{a_{n}}_{\;b_{n}}
=\varepsilon_{a_{1}\ldots a_{n}}
t^{a_{1}}_{\;c_{1}}
\ldots
t^{a_{n}}_{\;c_{n}}
P^{c_{1}\ldots c_{n}}_{\;b_{1}\ldots b_{n}}
= {\cal D} \varepsilon_{b_{1}\ldots b_{n}}.
$$
This proves (\ref{detrel}) and also
implies that
${\cal D}$ is grouplike:
$$\begin{array}{rcl}
\Delta {\cal D} &=&
\nu^{-1}
\varepsilon_{a_{1}\ldots a_{n}}
t^{a_{1}}_{\;c_{1}}
\ldots
t^{a_{n}}_{\;c_{n}}
\otimes
t^{c_{1}}_{\;b_{1}}
\ldots
t^{c_{n}}_{\;b_{n}}
\varepsilon^{b_{n}\ldots b_{1}}\\
&=&
\nu^{-1}
{\cal D}
\varepsilon_{c_{1}\ldots c_{n}}
\otimes
t^{c_{1}}_{\;b_{1}}
\ldots
t^{c_{n}}_{\;b_{n}}
\varepsilon^{b_{n}\ldots b_{1}}\\
&=& {\cal D} \otimes {\cal D}.
\end{array}
$$
Moreover, (\ref{detrel})
implies for the square of the
q-determinant on the
quotient $A$
$$
\begin{array}{rcl}
1&=&
\nu^{-1}
\varepsilon^{b_{1}\ldots b_{n}}
\varepsilon_{b_{n}\ldots b_{1}}\\
&=&\nu^{-1}\varepsilon_{a_{1}\ldots a_{n}}
g^{a_{1}b_{1}}
\ldots
g^{a_{n}b_{n}}
\varepsilon_{b_{n}\ldots b_{1}}\\
&=&\nu^{-1}
\varepsilon_{c_{1}\ldots c_{n}}
t^{c_{1}}_{\;a_{1}}
\ldots
t^{c_{n}}_{\;a_{n}}
g^{a_{1}b_{1}}
\ldots
g^{a_{n}b_{n}}
t^{d_{n}}_{\;b_{n}}
\ldots
t^{d_{1}}_{\;b_{1}}
\varepsilon_{d_{n}\ldots d_{1}}\\
&=&\nu^{-1}{\cal D}
\varepsilon_{a_{1}\ldots a_{n}}
g^{a_{1}b_{1}}
\ldots
g^{a_{n}b_{n}}
t^{d_{n}}_{\;b_{n}}
\ldots
t^{d_{1}}_{\;b_{1}}
\varepsilon_{d_{n}\ldots d_{1}}\\
&=&{\cal D}^{2},
\end{array}
$$
where we used  (\ref{epup}) and (\ref{metrel}).

The application of the
fundamental representation
$\rho_{-}$ from (\ref{fundrep}) to equation
(\ref{detrel}) yields
$$
\varepsilon_{a_{1}\ldots a_{n}}
R^{-1 ia_{1}}_{\;\;\;\;\;\;c_{1}b_{1}}
R^{-1 c_{1}a_{2}}_{\;\;\;\;\;\;c_{2}b_{2}}
\ldots
R^{-1 c_{n-1}a_{n}}_{\;\;\;\;\;\;m \; b_{n}}
=
\rho_{-}({\cal D})
\varepsilon_{b_{1}\ldots b_{n}}
\delta^{i}_{\;m},$$
and since ${\cal D}^{2}=1$ on the quotient $A$,
one
immediately finds $\rho_{-}({\cal D}) =1$.
Together with (\ref{qgrpnorm}) -- here
used in the form
$R^{-1 eb}_{\;\;\;\;\;\;fd}g^{ad}=
R^{ea}_{\;fd}g^{db}$ --
this
 relation also implies
$$
R^{ib_{1}}_{\;c_{1}a_{1}}
R^{c_{1}b_{2}}_{\;c_{2}a_{2}}
\ldots
R^{c_{n-1}b_{n}}_{\;\;m\;a_{n}}
\varepsilon^{a_{1}\ldots a_{n}}
=
\varepsilon^{b_{1}\ldots b_{n}}
\delta^{i}_{\;m}
$$
and finally with (\ref{arrels}) that
 ${\cal D}$ is central in $A(R)$:
$$
\begin{array}{rcl}
t^{i}_{\;j}\nu {\cal D}
&=&
t^{i}_{\;j}
\varepsilon_{a_{1}\ldots a_{n}}
t^{a_{1}}_{\;b_{1}}
\ldots
t^{a_{n}}_{\;b_{n}}
\varepsilon^{b_{n}\ldots b_{1}}\\
&=&
\varepsilon_{a_{1}\ldots a_{n}}
R^{-1 ia_{1}}_{\;\;\;\;\;\;c_{1}d_{1}}
\ldots
R^{-1 c_{n-1}a_{n}}_{\;\;\;\;\;\;m d_{n}}
t^{d_{1}}_{\;e_{1}}
\ldots
t^{d_{n}}_{\;e_{n}}
R^{ie_{1}}_{\;f_{1}b_{1}}
\ldots
R^{f_{n-1}e_{n}}_{\;\;m\;b_{n}}
\varepsilon^{b_{n}\ldots b_{1}}\\
&=& \nu {\cal D}  t^{i}_{\;j}
\end{array}
$$
This proves the proposition.
\hfill  {\bf q.e.d.}

This construction reproduces all known
quantum determinants,  and has the advantage
that it works quite generally.
As a new example, one can
construct a quantum
determinant for the
quantum Lorentz group ${\cal L}_{q}$
\cite{DAMTP/93-45}.
This algebra is
given as a quotient of $A({\bf R}_{L})$
by a metric
relation of the form (\ref{metrel})
which satisfies
(\ref{qgrpnorm}).
In terms
of the standard $SU_{q}(2)$ R-matrix
$$
R =  \left( \begin{array}{cccc} q &0 &0 &0\\
0 & 1
&q-q^{-1} & 0\\
0 &0& 1& 0\\0&0&0&q\end{array}\right),
\;\;\;\;\;q\in\rn,
$$
the R-matrix ${\bf R}_{L}$ is given by
$$ {\bf R}^{\;\;AB}_{L\;\;CD}=
R^{dk}_{\;lb}
R^{b^{'}l}_{\;ma}
R^{a^{'}m}_{\;n c^{'}}
\widetilde{R}^{nc}_{\;d^{'}k}
$$
where we use multi-indices
$A=(aa^{'})$ and
$\widetilde{R}$ is defined as
$((R^{t_{2}})^{-1})^{t_{2}}$
where $t_{2}$ denotes transposition
in the second tensor
component.
In an earlier paper \cite{DAMTP/94-10}
we already solved
the equations (\ref{quanti})
for this special case. The
one-dimensional space of
solutions has a
basis vector $\varepsilon_{ABCD}$ the
non-zero
entries of which are
$$\begin{array}{lllll}
\epsilon_{1234}=1&
\epsilon_{1243}=-q^{-2}&
\epsilon_{1324}=-1&
\epsilon_{1342}= q^{2}&
\epsilon_{1414}= 1-q^{2}\\
\epsilon_{1423}=1&
\epsilon_{1432}= -1&
\epsilon_{1444}= 1-q^{-2}&
\epsilon_{2134}= -1&
\epsilon_{2143}= q^{-2}\\
\epsilon_{2314}= 1&
\epsilon_{2341}= -1&
\epsilon_{2413}= -q^{-2}&
\epsilon_{2431}=q^{-2}&
\epsilon_{2434}= q^{-2}-1\\
\epsilon_{3124}= 1&
\epsilon_{3142}= -q^{2}&
\epsilon_{3214}= -1&
\epsilon_{3241}= 1 &
\epsilon_{3412}= q^{2}\\
\epsilon_{3421}= -q^{2} &
\epsilon_{3424}= 1-q^{2}&
\epsilon_{4123}= -1&
\epsilon_{4132}= 1&
\epsilon_{4141}=q^{2}-1\\
\epsilon_{4144}= q^{-2}-1&
\epsilon_{4213}= 1&
\epsilon_{4231}= -1&
\epsilon_{4243}= 1-q^{-2}&
\epsilon_{4312}= -1\\
\epsilon_{4321}= 1&
\epsilon_{4342}= q^{2}-1&
\epsilon_{4414}= 1-q^{-2} &
\epsilon_{4441}= q^{-2}-1
\end{array}$$
The normalisation factor $\nu$ is given by
$$\nu=2q^{-2}(1+q^{2}+q^{4})(1+q^{2})^{2}.$$
Since $\cal D$ is central and grouplike,
one can
divide ${\cal L}_{q}$ by the ideal generated
by the relation
${\cal D}=1$
and obtains a q-deformation of
$SO(3,1)$.

\end{document}